\documentclass[onecolumn,floatfix,superscriptaddress,a4paper,
               showpacs,showkeys,nofootinbib,preprint]{revtex4}
\usepackage{epsfig}
\usepackage{latexsym}
\usepackage{xspace}
\usepackage{hyperref}
\usepackage[latin2]{inputenc}
\usepackage{indentfirst}
\usepackage{enumerate}

\usepackage{amsmath}
\usepackage{amssymb}
\usepackage[english]{babel}
\usepackage{url}

\newcommand{\eq}[1]{\begin{align} #1 \end{align}}

\begin{document}

\title{Hadron Resonance Gas Equation of State      \\ from  Lattice QCD}
\author{V. Vovchenko}
\affiliation{
Taras Shevchenko National University of Kiev, Kiev, Ukraine}
\affiliation{
Frankfurt Institute for Advanced Studies, Johann Wolfgang Goethe University, Frankfurt, Germany}
\affiliation{
GSI Helmholtzzentrum f\"ur Schwerionenforschung GmbH, Darmstadt, Germany}
\author{D. V. Anchishkin}
\affiliation{
Bogolyubov Institute for Theoretical Physics, Kiev, Ukraine}
\affiliation{
Taras Shevchenko National University of Kiev, Kiev, Ukraine}
\author{M. I. Gorenstein}
\affiliation{
Bogolyubov Institute for Theoretical Physics, Kiev, Ukraine}
\affiliation{
Frankfurt Institute for Advanced Studies, Johann Wolfgang Goethe University, Frankfurt, Germany}

\date{\today}


\begin{abstract}
The Monte Carlo results in lattice QCD for the pressure and energy density
at  small temperature $T < 155$~MeV and zero baryonic chemical potential are
analyzed within the hadron resonance gas model.
Two extensions of the ideal hadron resonance gas
are considered: the excluded volume model which describes a repulsion
of hadrons at short distances and Hagedorn model with the exponential mass spectrum.
Considering both of these models {\it one by one} we do not find
the conclusive evidences in favor of {\it any of them}.
The controversial results appear because of rather different sensitivities
of the pressure and energy density to both excluded volume and Hagedorn mass spectrum
effects. On the other hand, we have found a clear
evidence for a {\it simultaneous} presence of {\it both of them}.  They lead to
rather essential contributions: suppression effects for
thermodynamical functions of the hadron resonance gas due to the excluded
volume effects and  enhancement due to the Hagedorn mass spectrum.

\end{abstract}

\maketitle

\section{Introduction}
\begin{figure}[ht]
\centering
\includegraphics[width=0.69\textwidth]{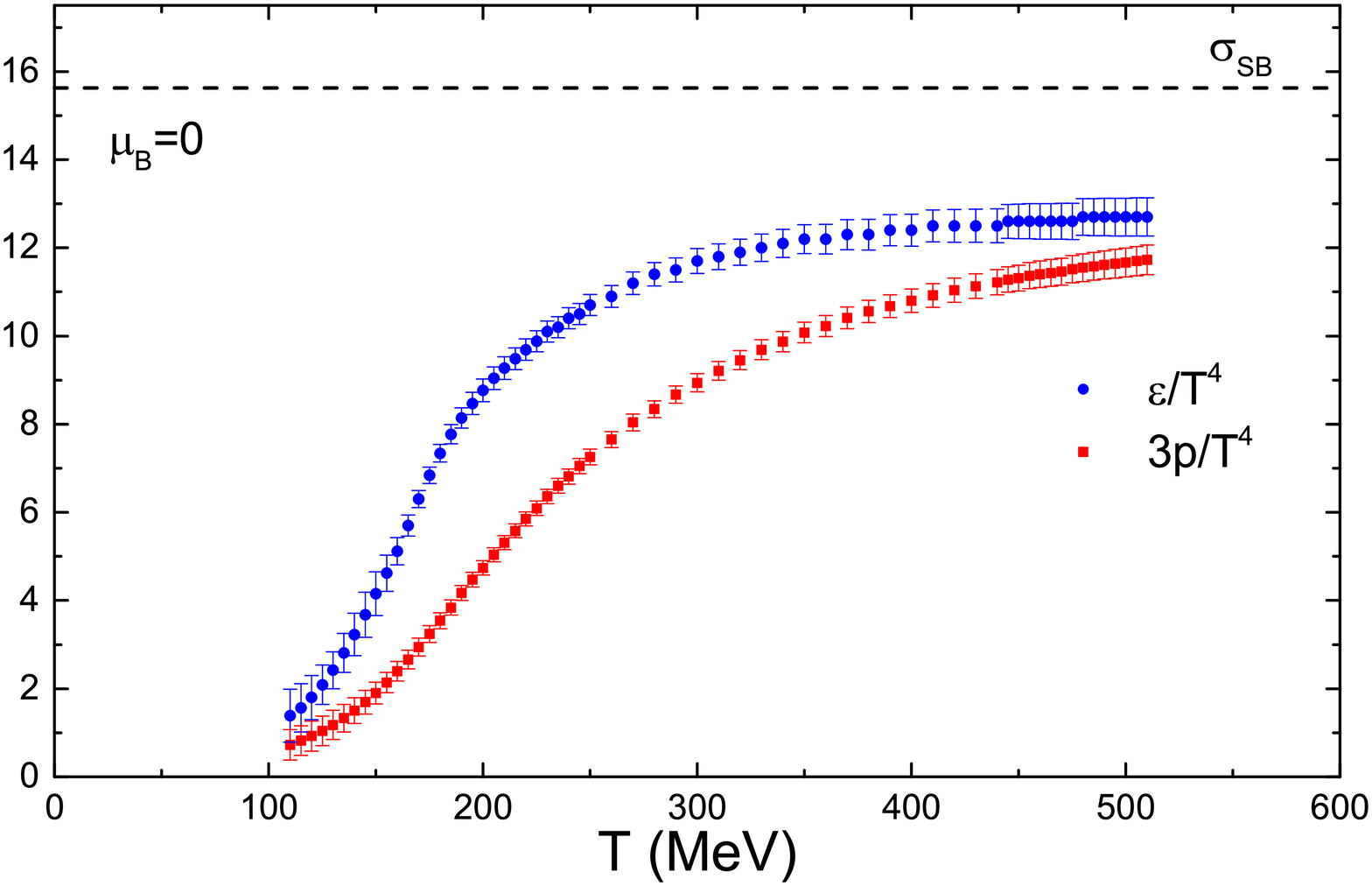}
\caption[]{The lattice results from Ref.~\cite{lattice-1c} for $3p/T^4$ (circles)
and $\varepsilon/T^4$ (squares) at zero baryonic chemical potential.
}\label{fig-latt}
\end{figure}

The Monte Carlo calculations in the lattice QCD at finite temperature $T$ (see, e.g.,
Refs.~\cite{lattice-1a,lattice-1b,lattice-1c,lattice-2} and references therein)
reveal two physical  phases of strongly interacting matter: hadron phase
at small $T$ and deconfined quark-gluon phase at high $T$.
In Fig.~\ref{fig-latt} the lattice results of Ref.~\cite{lattice-1c} for the
pressure and energy density ($3p/T^4$ and $\varepsilon/T^4$)
obtained at zero baryonic chemical potential in QCD with 2+1 quark flavors and
extrapolated to the thermodynamical and continuum limits are shown as functions
of temperature $T$.

From Fig.~\ref{fig-latt} one observes  a steep increase of thermodynamical
quantities near the crossover temperature $T_c$. This temperature is estimated
in the range of 150-160~MeV.  The values of $3p/T^4$ and $\varepsilon/T^4$
in  the deconfined quark-gluon phase approach slowly from below the Stefan-Boltzmann
limit $3p_{SB}/T^4=\varepsilon_{SB}/T^4=\sigma_{SB}$, which equals to
$\sigma_{SB}=19\pi^2/12\cong 15.6$ in the 3-flavor QCD.
At $T<T_c$ the confined hadron phase emerges.
In the present paper the lattice data \cite{lattice-1c}  will
be used to constrain an equation of state of the hadronic matter.

A description of hadron multiplicities in high-energy nucleus-nucleus collisions
shows a surprisingly good agreement between the  results
of the hadron resonance gas (HRG) model
(see, e.g.,
Refs.~\cite{Id-HRG-1,Id-HRG-2,Id-HRG-3,Id-HRG-4,Id-HRG-5,Id-HRG-6})
and the experimental data.
In most statistical model
formulations the ideal HRG (Id-HRG) is used. It
is argued  that a presence of all known resonance states in
the thermal system takes into account
attractive interactions between hadrons \cite{DMB}.

Two extensions of the Id-HRG model have been  widely discussed.
The first one is the excluded volume HRG (EV-HRG) model which
introduces the effects of hadron repulsions
at short distances.
One usually uses the van der Waals procedure
\cite{vdw-1,vdw-2} and substitutes a system volume $V$
by the available volume $V-\sum_i v_iN_i$, where $v_i$ is the volume
parameter for $i$th hadron species,
$N_i$ is the number of particles of $i$th type, and the sum is taken over all
types $i$ of hadrons and resonances.
Another example of attractive and
repulsive interactions between hadrons is the relativistic mean field theory
\cite{W-1} (see also the recent paper \cite{W-2} and references
therein).
Note that the EV-HRG model can be equivalently
formulated in terms of the mean-field (see Refs.~\cite{mf-1992,mf-1995,mf-2014}).
This makes it possible to incorporate other hadron interactions
within unified mean-field approach.

The second extension of the HRG model is an inclusion of the exponentially
increasing mass spectrum  $\rho(m)$ proposed by Hagedorn about 50
years ago \cite{hag,hag-ranft}.
These excited colorless states (named fireballs or strings) are considered as
a continuation of the resonance spectrum
at masses $m$ higher than $2$~GeV.

In the present paper we use the lattice data \cite{lattice-1c} at small
temperature $T<155$~MeV to confirm a presence of the excluded volume effects and effects of the
Hagedorn mass spectrum.
The HRG model had been used for comparison with the
lattice data in the hadronic sector~\cite{HRG-1-latt,HRG-2-latt,HRG-3-latt}.
There were
as well
several publications, where the EV-HRG model (see, e.g.,
Refs.~\cite{EV-1-latt, EV-2-latt}) or the Hagedorn mass spectrum
(see, e.g., Refs.~\cite{Hag-1,Hag-2,Hag-3}) were confronted with the lattice
data.
Our analysis extends these previous attempts to the case when  both these
physical effects
are treated simultaneously.
Included together they improve essentially an agreement of the HRG model with
the lattice results, while treated separately none of them can be clearly
established.

The paper is organized as follows.  In Sec.~\ref{sec-HRG} the grand canonical ensemble formulation
of the Id-HRG and EV-HRG are considered. In Sec.~\ref{sec-hag} the EV-HRG model is extended
by inclusion of the Hagedorn mass spectrum. A summary in Sec.~\ref{sec-sum}
closes the article.

\section{Hadron resonance gas}
\label{sec-HRG}

\subsection{Ideal Hadron Resonance Gas}
\label{sec-Id-HRG}
In the grand canonical ensemble the pressure and energy density of the Id-HRG
are given by
\begin{eqnarray}
\label{p-id}
p^{\rm id}(T,\mu) &=& \sum_i p_i^{\rm id}(T,\mu_i)
\nonumber \\
&=& \sum_i\frac{d_i}{6\pi^2}\int d m\, f_i(m)\int_0^{\infty} \frac{k^4dk}{\sqrt{k^2+m^2}}
\left[ \exp\left(\frac{\sqrt{k^2+m^2} - \mu_i}{T}\right)+\eta_i\right]^{-1}~,
\\
\varepsilon^{\rm id}(T,\mu) &=& \sum_i \varepsilon_i^{\rm id}(T,\mu_i)
\nonumber \\
&=& \sum_i\frac{d_i}{2\pi^2}\int d m\, f_i(m)\int_0^{\infty}k^2dk\,\sqrt{k^2+m^2}
\left[ \exp\left(\frac{\sqrt{k^2+m^2}-\mu_i}{T}\right)~+~\eta_i\right]^{-1}~,
\label{e-id}
\end{eqnarray}
where $d_i$ is the spin degeneracy of $i$th particle and the normalized
function $f_i(m)$
takes into account the Breit-Wigner shape of resonance with finite width
$\Gamma_i$ around their average mass $m_i$, for the stable hadrons,
$f_i(m)=\delta(m-m_i)$.
The sum over $i$ in Eqs.~\eqref{p-id} and \eqref{e-id} is taken over all
non-strange and strange hadrons that are listed in Particle Data Tables
~\cite{pdg}.
This includes mesons up to $f_2(2340)$
and (anti-)baryons up to $N(2600)$.
We also note that in these equations
$\eta_i =-1$ and $\eta_i = 1$ for bosons
and fermions, respectively, while $\eta = 0$ corresponds to the Boltzmann approximation.
The chemical potential for $i$th hadron is given by
\begin{equation}
\mu_i\ =\ b_i\,\mu_B\, +\, s_i\,\mu_S\, +\, q_i\,\mu_Q
\label{eq:mui}
\end{equation}
with $b_i = 0,\, \pm 1$, $s_i = 0,\, \pm 1,\, \pm 2,\, \pm 3$, and
$q_i = 0,\, \pm 1,\, \pm 2$
being the corresponding baryonic number, strangeness, and electric charge of
$i$th  hadron.
The notation $\mu$
will be used to denote all chemical potentials, $\mu\equiv (\mu_B,\mu_S,\mu_Q)$.

\subsection{Excluded Volume Hadron Resonance Gas}
In this subsection a role of the repulsive interactions is considered within
the EV-HRG model.
The van der Waals excluded volume procedure corresponds to a substitution of
the system volume $V$ by the available volume $V_{av}$,
\begin{equation}
V \rightarrow  V_{\rm av}\, =\, V - \sum_i v_i N_i~,
\label{vdw}
\end{equation}
where $N_i$ is the particle number,  $v_i = 4\cdot (4\pi r_i^3/3)$
is the excluded volume parameter with $r_i$ being the corresponding hard
sphere radius of particle $i$, and the sum is taken over all hadrons and resonances.
This result, in particular, the presence of a factor of 4 in the expression for
$v_i$, can be rigorously obtained for a low density gas of particles of a single
type (see, e.g., Ref. \cite{LL}).
In the grand canonical ensemble, the substitution (\ref{vdw}) leads to a
transcendental equation for the EV-HRG pressure \cite{vdw-2}:
\eq{\label{p-ev}
p^{\rm ev}(T,\mu)=\sum_i p_i^{\rm id}(T,\tilde{\mu_i})~;~~~~\tilde{\mu_i}=\mu_i- v_i p^{\rm ev}~,
}
and the energy density is calculated as
\eq{\label{e-ev}
\varepsilon^{\rm ev}(T,\mu) = \frac{\sum_i\varepsilon_i^{\rm id}(T,\tilde{\mu_i})}{1+\sum_j v_j
n_j^{\rm id}(T,\tilde{\mu_j})}~,
}
where $n_i^{\rm id}$ is the ideal-gas particle number density of $i$th hadron species,
\eq{\label{n-id}
n^{\rm id}_i(T,\mu_i)\, =\, \frac{d_i}{2\pi^2}\int dm f_i(m)\int_0^{\infty}k^2dk\,
\Big[ \exp\Big(\frac{\sqrt{k^2+m^2}-\mu_i}{T}\Big)~+~\eta_i\Big]^{-1}~.
}

In what follows we restrict our consideration to the case of equal volume parameters
$v_i$ for all hadrons and resonances, $v_i= v\equiv 16\pi r^3/3$.
The Boltzmann approximation $\eta_i=0$ in Eqs.~(\ref{p-id},\ref{e-id},\ref{n-id})
simplifies Eqs.~(\ref{p-ev}) and (\ref{e-ev}):
\eq{\label{p-ev-1}
p^{\rm ev}(T,\mu)&= \kappa^{\rm ev}\,
p^{\rm id}(T,\mu)~=~\kappa^{\rm ev}\,T\,
n^{\rm id}(T,\mu)~,\\
\varepsilon^{\rm ev}(T,\mu) &= \frac{\kappa^{\rm ev}\,\varepsilon^{\rm id}(T,\mu)}
{1~+~v\,\kappa^{\rm ev}\,n^{\rm id}(T,\mu)}~,
\label{e-ev-1}
}
where the excluded volume suppression factor $\kappa_{\rm ev}$ and the total
particle number density $n^{\rm id}$ in the Id-HRG are introduced as
\begin{equation}
\kappa^{\rm ev}~\equiv~\exp\left(-~\frac{v\,p^{\rm ev}}{T}\right)~,~~~~~
n^{\rm id}(T,\mu)~\equiv~\sum_i n_i^{\rm id}(T,\mu_i)~.
\label{kappa}
\end{equation}
Expressions (\ref{p-ev-1}-\ref{kappa}) can be also obtained
in the framework of thermodynamically self-consistent
mean-field theory (see Sec.~V in Ref. \cite{mf-2014}).
This approach gives a sequential treating of the problem when one
can examine various different mean fields that mimic the repulsive and
attractive interactions
(for details see Refs.~\cite{mf-1992, mf-1995, mf-2014}).

The EV-HRG
was used to fit
the data on hadron multiplicities in Ref.~\cite{EV-HRG-1} with values of $r$ in
the region of 0.2-0.8~fm.
A numerical value of the hard-core radius was estimated as $r=0.3$~fm in
Ref.~\cite{EV-HRG-2}.
Note that if radii of all hadrons are assumed to be the same, the chemical
freeze-out parameters, temperature and baryon chemical potential, fitted to
data on hadron multiplicities, are identical to those obtained within the
Id-HRG model.
Indeed,  the particle number ratios are not sensitive to the numerical value
of $r$.
Hence, in order to establish a presence of non-zero hard-core hadron radii,
the independent measurements of a total system volume are needed.
On the other hand, it was shown that
the particle number fluctuations depend straightforwardly on the hard-core
hadron radius \cite{EV-fluc-1,EV-fluc-2}.
Thus, an interpretation of these data within the EV-HRG opens the way to
estimate the value of $r$ from the data.
This is, however, not an easy task as there are many other effects which
influence the particle number fluctuations.

\subsection{Id-HRG and EV-HRG versus the Lattice Data}

\begin{figure}[ht]
\centering
\includegraphics[width=0.49\textwidth]{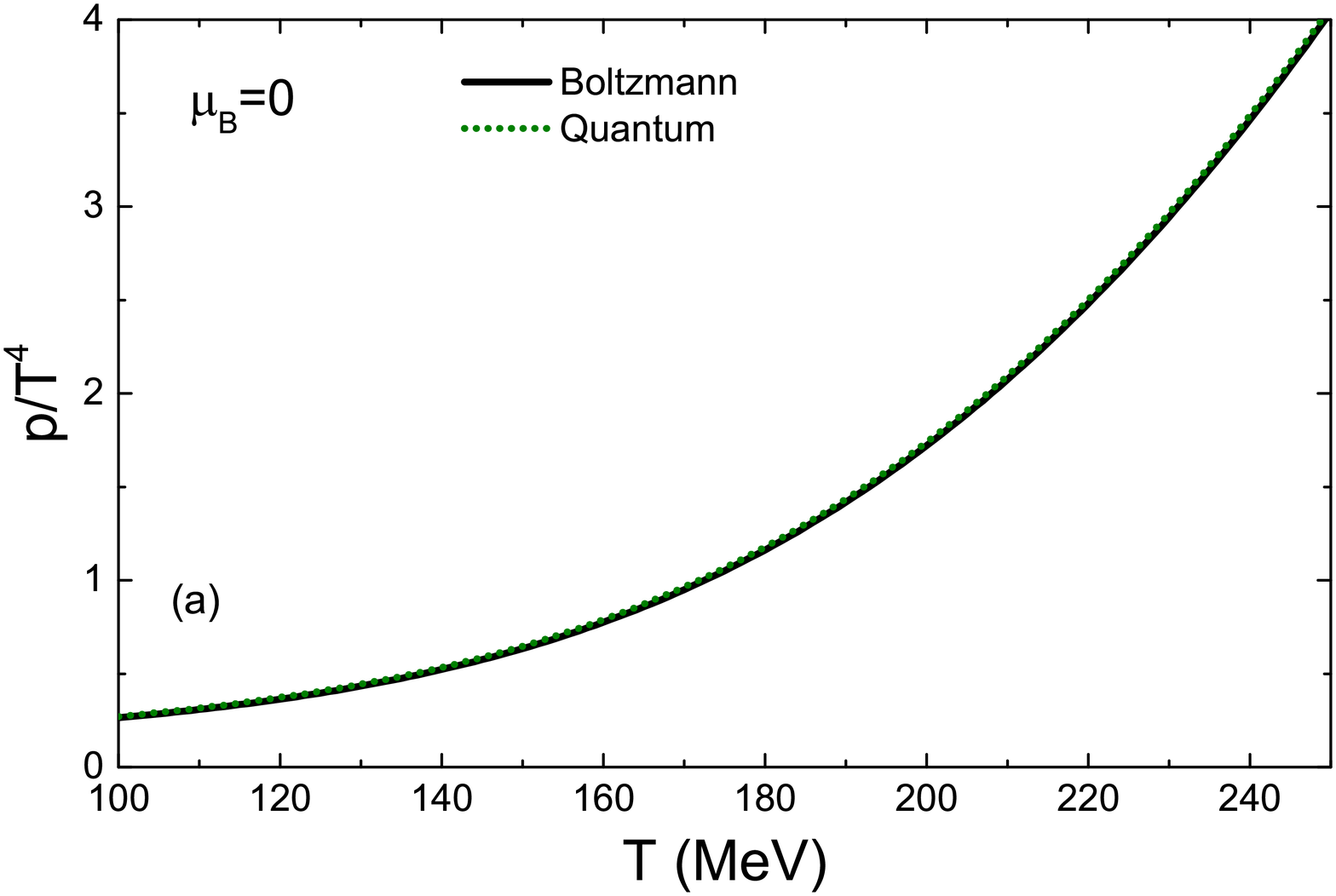}
\includegraphics[width=0.49\textwidth]{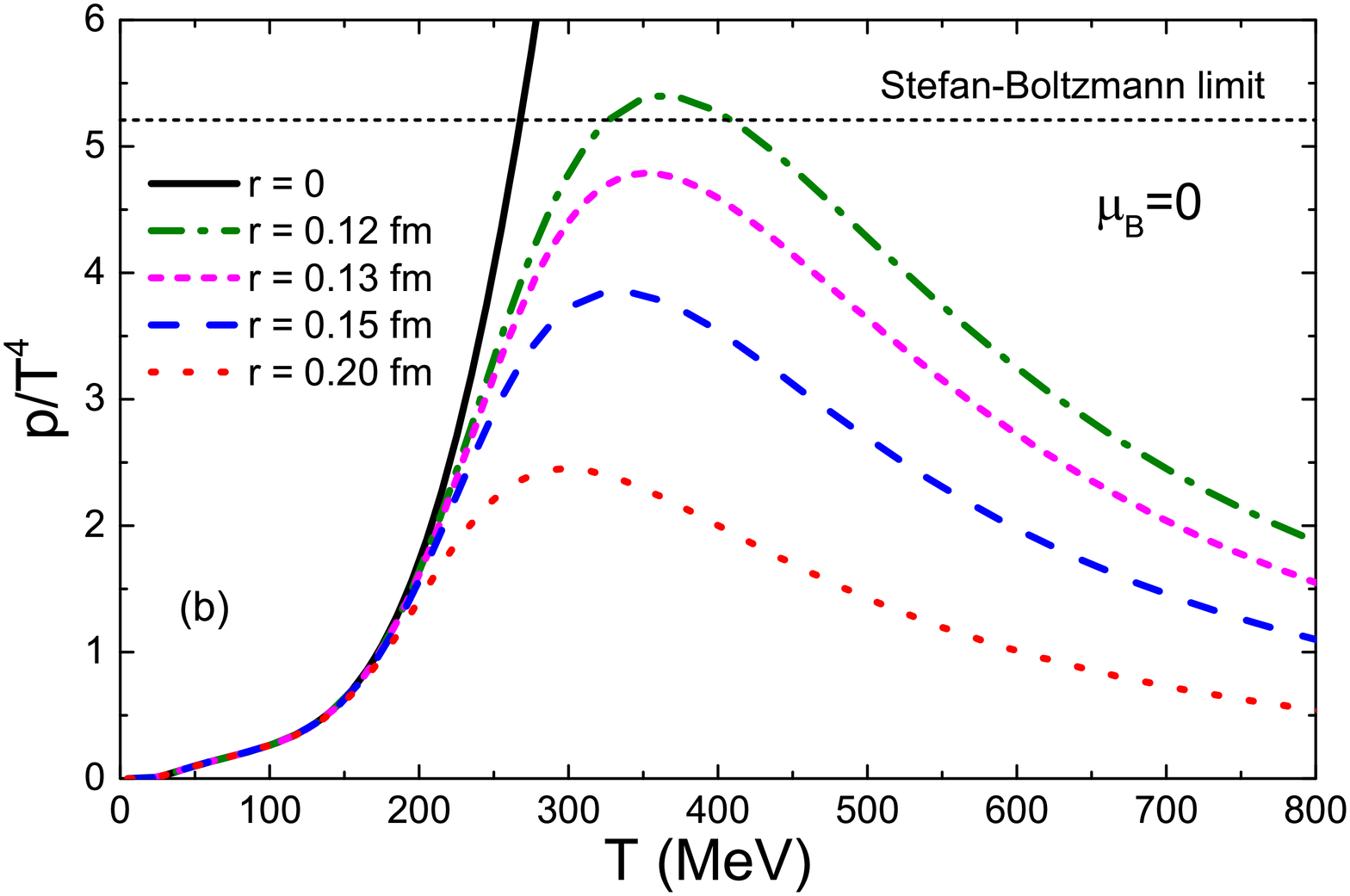}
\caption[]{
(a): The Id-HRG pressure, $p^{\rm id}(T)/T^4$, is shown as a
function of temperature at $\mu=0$ by the dotted line.
The Boltzmann approximation
$\eta_i=0$ is shown by the solid line.
(b): The Id-HRG pressure and EV-HRG pressure functions for several different
values of hard-core radius $r$ are presented.
The Stefan-Boltzmann limit for the deconfined quark-gluon phase,
$p_{SB}/T^4=\sigma_{SB}/3\cong 5.2$ is indicated by the horizontal dotted line.
}
\label{fig-id}
\end{figure}

In Fig.~\ref{fig-id} (a) the Id-HRG pressure (\ref{p-id}) divided
 by
$T^4$ is shown as a function of temperature for the case of zero chemical potentials,
\eq{\label{mu0}
\mu_B=\mu_S=\mu_Q=0~ .
}
Equation (\ref{mu0}) corresponds to zero values of all conserved charges,
baryonic number, strangeness, and electric charge, of the strongly interacting
matter.
This is approximately valid for the matter created at the Large Hadron Collider
(LHC) of the European Organization for Nuclear Research (CERN).
As it is seen from Fig.~\ref{fig-id} (a)
the Boltzmann approximation $\eta_i=0$ (solid line)
gives a very accurate evaluation
of $p^{\rm id}(T)/T^4$ calculated with the quantum statistics
(dashed line). In fact, a difference between the solid and dashed lines
is hardly seen in Fig.~\ref{fig-id} (a). The Boltzmann approximation
will be thus
adopted for our further analysis.
Note that a shift of the chemical potential according to Eq.~(\ref{p-ev})
makes the Boltzmann approximation in the EV-HRG
even more accurate than in the case of the Id-HRG model.

\begin{figure}[ht]
\centering
\includegraphics[width=0.49\textwidth]{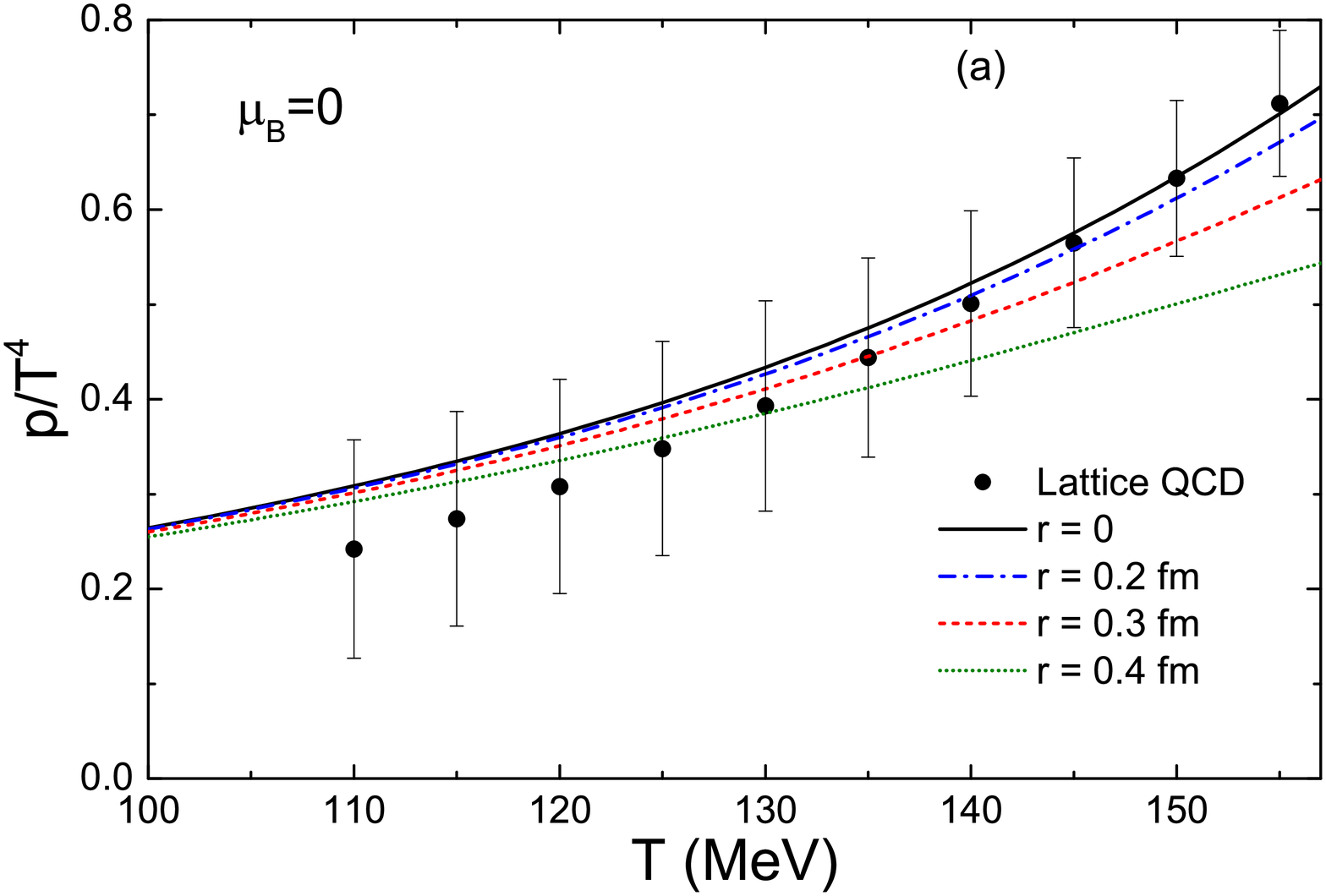}
\includegraphics[width=0.49\textwidth]{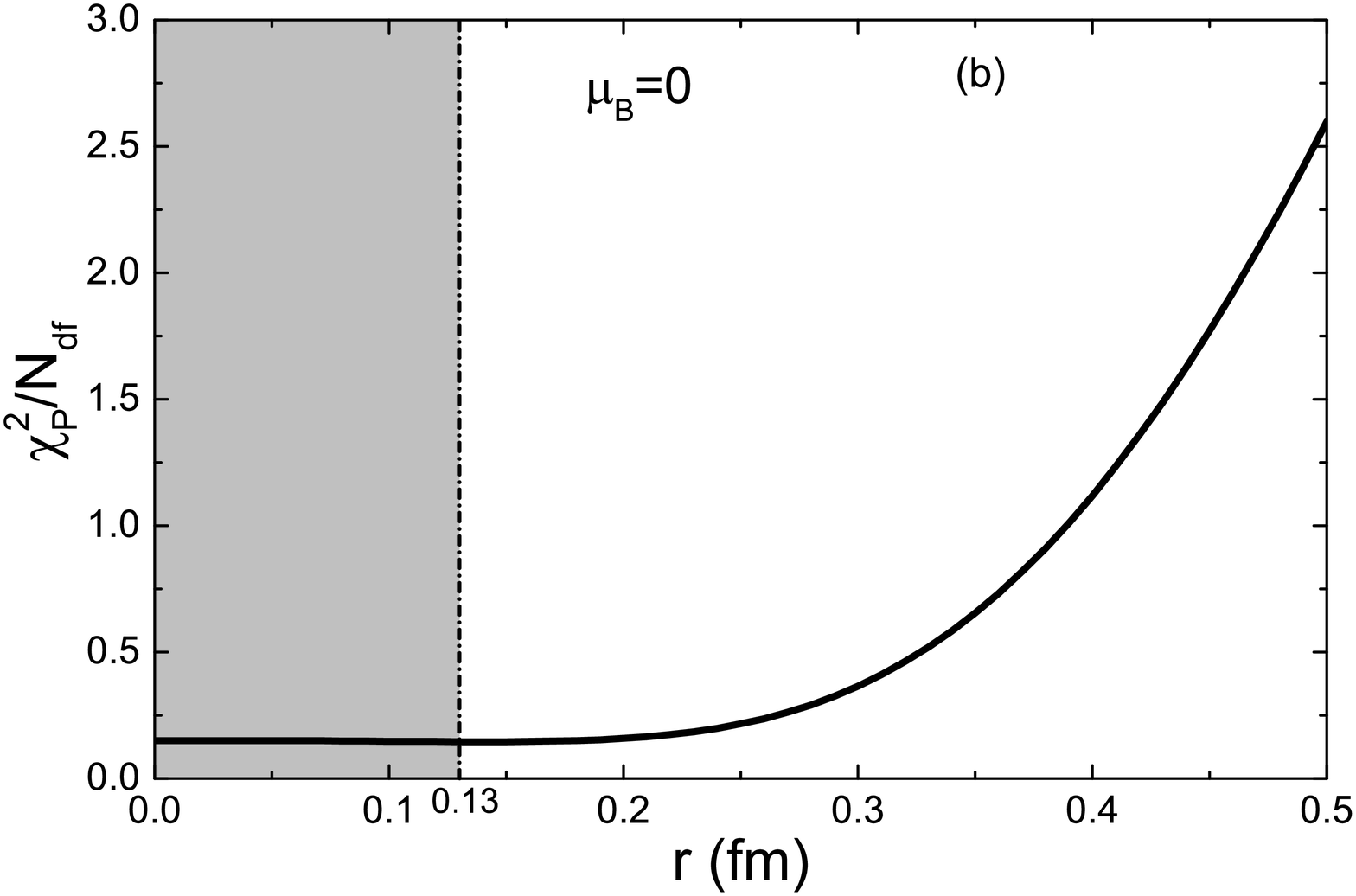}
\includegraphics[width=0.49\textwidth]{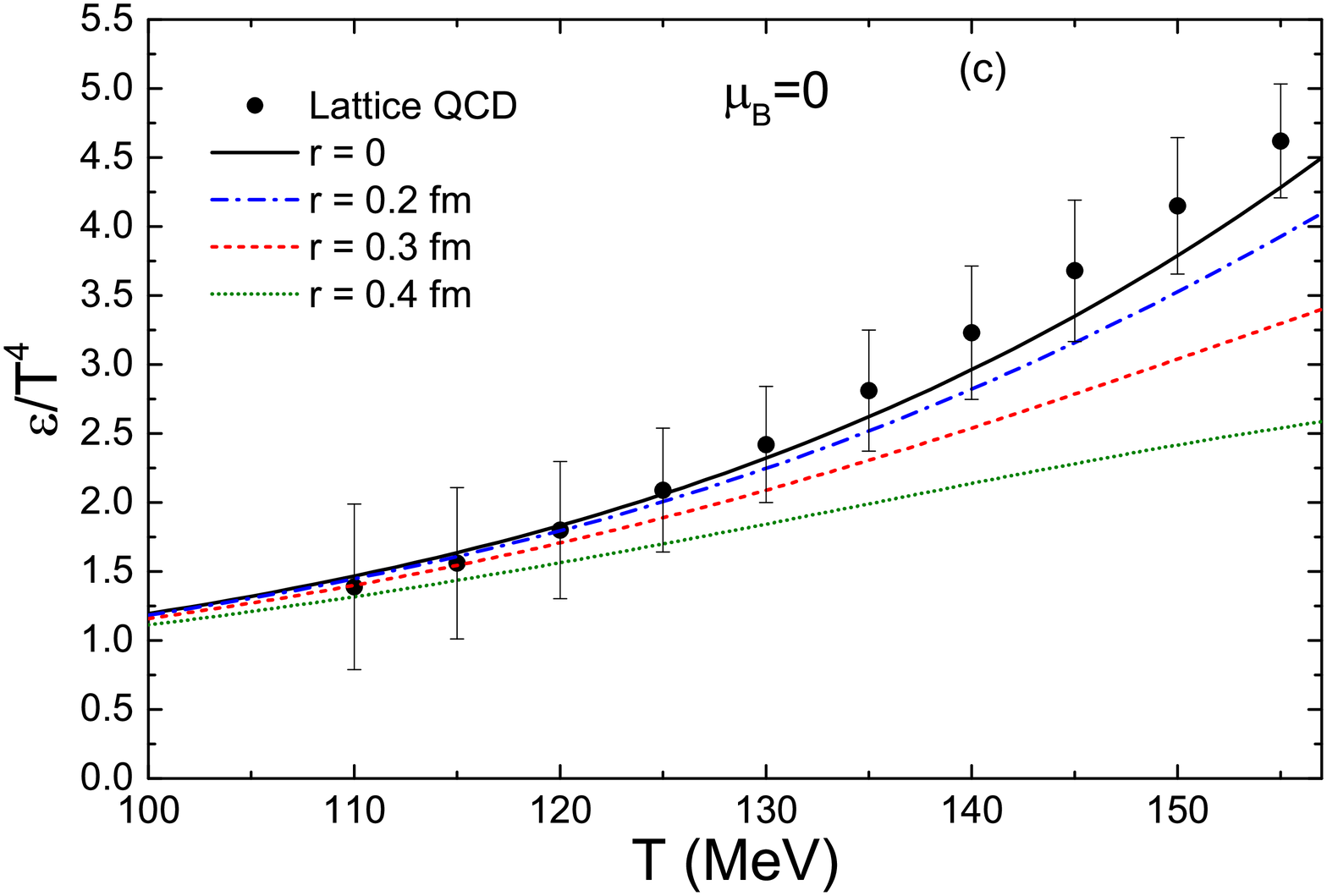}
\includegraphics[width=0.49\textwidth]{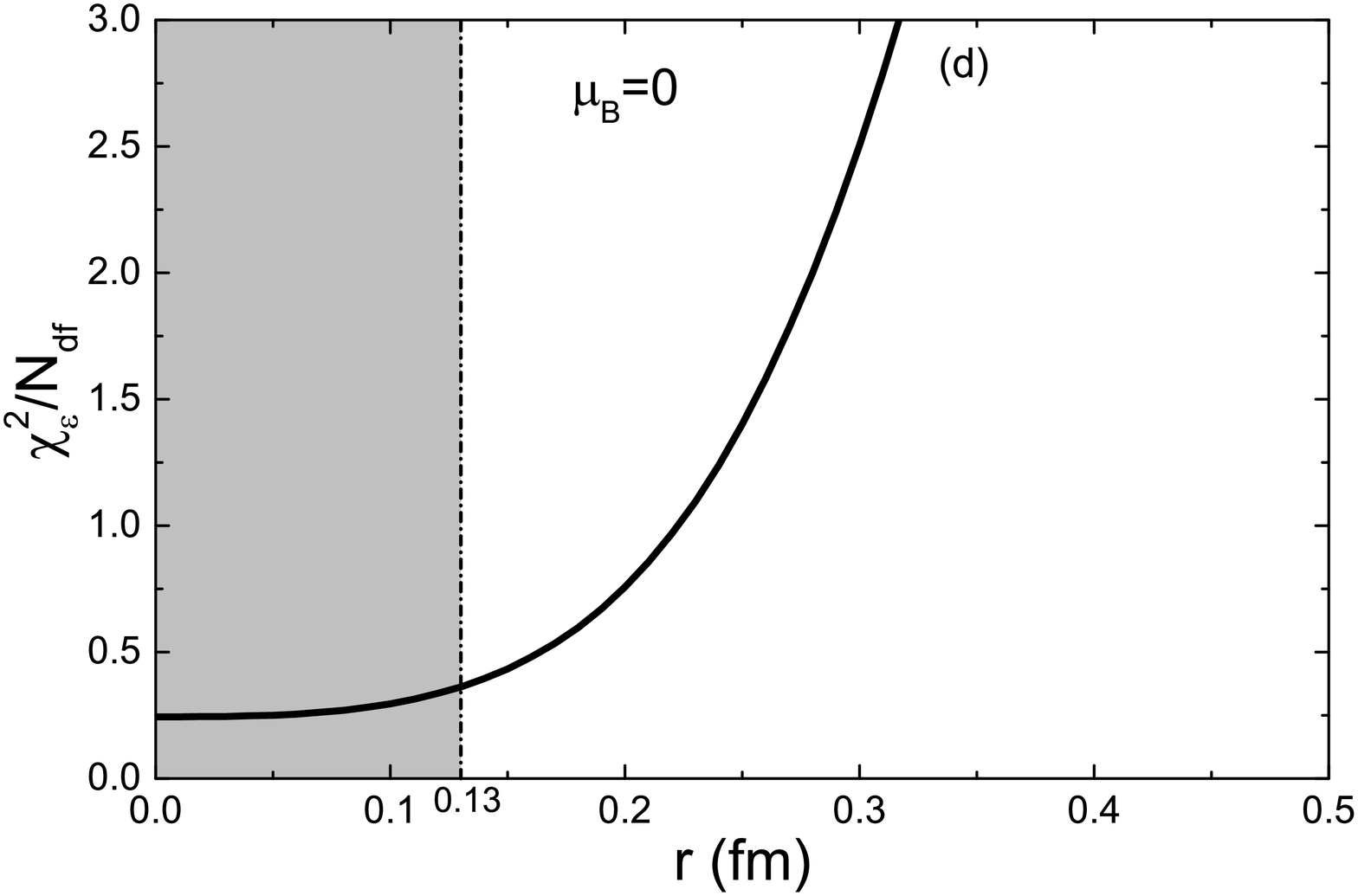}
\caption[]{
The results of EV-HRG model for different values of $r$ are compared to the
lattice data for $p/T^4$ (a) and $\varepsilon/T^4$ (c).
The values of $\chi^2_p/N_{\rm df}$  and $\chi^2_{\varepsilon}/N_{\rm df}$
are shown as functions of $r$ in (b) and (d), respectively,
and the shaded grey area corresponds to
$r\leq 0.13$~fm.
}
\label{fig-ev}
\end{figure}

In Fig.~\ref{fig-id} (b) the Id-HRG pressure (\ref{p-id}) is compared with
EV-HRG pressures (\ref{p-ev-1}) at several values of the hard-core radius $r$.
Note that the Id-HRG model shows a strong increase of $p^{id}(T)/T^4$ at high $T$
which exceeds the Stefan-Boltzmann pressure of the deconfined quarks and gluons,
$p_{SB}(T)/T^4=\sigma_{SB}/3\cong 5.2$.
Therefore, according to the Gibbs criterium the point-like hadrons would always
be the dominant phase at high temperatures \cite{vdw-3} due to the large number
of different types of mesons and baryons.
This
feature of the Id-HRG equation of state contradicts to the lattice QCD
results, hence it shows a shortcoming of the model based on the concept of the
point-like particles.
Just the excluded volume effects ensure a transition
from a gas of hadrons and resonances to the quark-gluon plasma. One needs
therefore the
EV-HRG equation of state for the hydrodynamic model calculations of nucleus-nucleus collisions
(see, e.g., Refs.~\cite{hyd-1,hyd-2,hyd-3}).
For any finite particle volume, i.e. $r>0$, the behavior of pressure is found as
$p^{\rm ev}(T)/T^4\cong (vT^3)^{-1}\rightarrow 0$
at $T\rightarrow \infty$.
However, as seen from Fig.~\ref{fig-id} (b), a more rigid restriction,
$r\ge 0.13$~fm, is needed to guarantee
that
$p^{\rm ev}(T)/T^4<\sigma_{SB}/3$ at all $T>T_c$.

In Fig.~\ref{fig-ev} (a), the EV-HRG results for $p^{\rm ev}/T^4$ are compared to the lattice
data $p^{\rm lat}/T^4$ \cite{lattice-1c}
at $T<155$~MeV for several different values of hard-core radius $r$. In Fig.~\ref{fig-ev} (b), the value
of $\chi^2_p/N_{\rm df}$ at different $r$ is shown. This quantity is calculated as
\eq{\label{chi-p}
\chi^2_p/N_{\rm df} = \frac{1}{N_{\rm df}} \, \sum_{i=1}^{N}
\frac{[(p^{\rm ev}/T^4)_i-(p^{\rm lat}/T^4)_i]^2}{[\Delta(p^{\rm lat}/T^4)_i]^2}~,
}
where
$N_{\rm df}$ is number of points $N$ (equal 10 in our case) minus the number of fitting parameters
(one parameter $r$ in our fit).
A most essential part of uncertainties in the lattice
data is not statistical. The systematical uncertainties
dominate, and these uncertainties are significantly correlated.
In this case, the usage of
$\chi^2/N_{\rm df}$ criterion is not perfectly reasonable.
However, we still use this quantity as a way to quantify the deviations
of HRG calculations from the lattice data.

From Fig.~~\ref{fig-ev} (b) one observes that the lattice data
for $p^{\rm lat}/T^4$  are fitted well in a rather wide
range of hard-core radius of $r\lesssim0.4$~fm.
Therefore, the lattice data for the hadron pressure
are consistent with a presence of rather significant excluded volume effects
and suggest reasonable numerical values for the hard-core radius $r$.
However, a comparison of the EV-HRG
model with the lattice results
$\varepsilon^{\rm lat}/T^4$ shown in Fig.~\ref{fig-ev} (c)
does not indicate a presence of the excluded volume effects.
This is clear from Fig.~\ref{fig-ev} (d), where we show the value of
$\chi^2_{\varepsilon}/N_{\rm df}$ calculated as in Eq.~(\ref{chi-p}) but with
$\varepsilon/T^4$ instead of $p/T^4$. The value of $r=0$ corresponds to the best fit
of  $\varepsilon^{\rm lat}/T^4$. Let us, however, remind Fig.~\ref{fig-id} (b) which
shows that too small values
of the hard-core radius, $r\leq 0.13$~fm, look doubtful. For these small values of $r$
the EV HRG pressure becomes larger than the Stefan-Boltzmann limit for quarks and gluons.
The `forbidden' region of the hard-core radius, $r<0.13$~fm, is shown as
the grey area in Figs. \ref{fig-ev} (b), (d)
and \ref{fig-hag2} (a), (b).

Therefore, while a
reasonable $r$-value, e.g., $r=0.3$~fm, gives a good agreement, $\chi^2_p/N_{\rm df}\cong 0.4$,
with the $p^{\rm lat}/T^4$ lattice data, it also leads
to rather large value of $\chi^2_{\varepsilon}/N_{\rm df}\cong 2.5$ and, thus,
looks unreasonable for the fit of $\varepsilon^{\rm lat}/T^4$.

\section{Excluded Volume HRG with Hagedorn mass spectrum}
\label{sec-hag}

The analysis presented in the previous section gives no conclusive answer about
a presence of the excluded volume effects.
The value of $r=0.3$~fm in the EV-HRG model leads
to sizeable suppression
effects of the Id-HRG pressure and
to a good
agreement with the lattice data
$p^{\rm lat}/T^4$, whereas
the $\varepsilon^{\rm lat}/T^4$ data
prefer the value of $r\cong 0$ and is thus consistent with the Id-HRG model.
From our point of view, this observation may indicate a presence of additional
contributions to $p^{\rm ev}$ and $\varepsilon^{\rm ev}$ in the EV-HRG model.
These contributions should be small enough for the pressure and much larger for
the energy density.
We argue that massive Hagedorn states are the ideal candidates for this role.
Indeed, each heavy particle with $m\gg T$ gives its contribution $T$  to the
pressure, and  much larger contribution, $m+3T/2$, to the energy density.

For a further analysis we use the following parametrization for the Hagedorn
mass spectrum \cite{hag-ranft}:
\eq{\label{rho}
\rho(m)~=~C~\frac{\operatorname\theta(m-M_0)}{(m^{2}+m_0^2)^{a}}~\exp\left(\frac{m}{T_H}\right)~.
}
The spectrum (\ref{rho})
with the following parameters, $M_0=2$~GeV, $T_H=160$~MeV,
$m_0$=0.5~GeV, and $a=5/4$, will be used.
The parameter $C$ in (\ref{rho}) will be the only free parameter
in the following analysis. We have checked that another set of parameters,
e.g.,
the same set with $M_0=2.5$~GeV, and also
a set with $m_0=0$, $a=3/2$, and $M_0=3$~GeV, leads to very similar results.

Our final assumption concerns the proper volume for the Hagedorn states.
To avoid
additional free parameters
we adopt the same value of $v=16\pi r^3/3$
for all known hadrons and resonances as well as for the Hagedorn
states.
The EV-HRG model with the Hagedorn mass spectrum will be denoted as EV-HRG-H.
The pressure $p^{\rm H}(T)$ in EV-HRG-H  model is  given by the following
equation
\begin{equation}
p^{\rm H}=\exp\left(-\frac{vp^{\rm H}}{T}\right)\,T\,  \int dm  \int_0^{\infty}
\frac{k^2 dk}{2\pi^2}\,\exp\left(-\,\frac{\sqrt{m^2+k^2}}{T}\right)
 \left[\sum_{i}d_i f_i(m)+\rho(m)\right] ~.
\label{pHag}
\end{equation}
The expression (\ref{pHag})
can be equivalently rewritten similar to Eq.~(\ref{kappa}),
\begin{equation}
p^{\rm H}(T)~=~\kappa^{\rm H}\,p^{\rm id}_{\rm H}(T)~
=~\kappa^{\rm H}\,T\,n^{\rm{id}}_{\rm H}(T)~,
\label{n-H}
\end{equation}
where $\kappa^{\rm H}\equiv \exp(-vp^{\rm H}/T)$, and $p^{\rm H}_{\rm id}$ and
$n^{\rm H}_{\rm id}$ are, respectively, the expressions for the pressure and
total number density of all particles  (hadrons, resonances, and Hagedorn
excited states)  in the ideal HRG with the Haggedorn mass spectrum (Id-HRG-H),
i.e., with the Hagedorn mass spectrum
but without excluded volume effects.
One can easily calculate the energy density as
\begin{equation}
\varepsilon^{\rm H}(T)~=~ T\, \frac{dp^{\rm H}}{dT}~-~p^{\rm H}~=~
\frac{\kappa^{\rm H}\,\varepsilon^{\rm H}_{\rm id}}{1~+~v\,\kappa^{\rm H}\,n^{\rm H}_{\rm id}}~,
\label{e-hag}
\end{equation}
where
\begin{equation}
\varepsilon^{\rm H}_{\rm id}(T)~=~\int dm  \int_0^{\infty}
\frac{k^2 dk}{2\pi^2}\, \sqrt{m^2+k^2}\,\exp\left(-\,\frac{\sqrt{m^2+k^2}}{T}\right)
 \Big[\sum_{i}d_i f_i(m)+\rho(m)\Big] ~
\label{eq:hag-energy-density}
\end{equation}
denotes the energy density of the  Id-HRG-H model.

\begin{figure}[ht]
\centering
\includegraphics[width=0.49\textwidth]{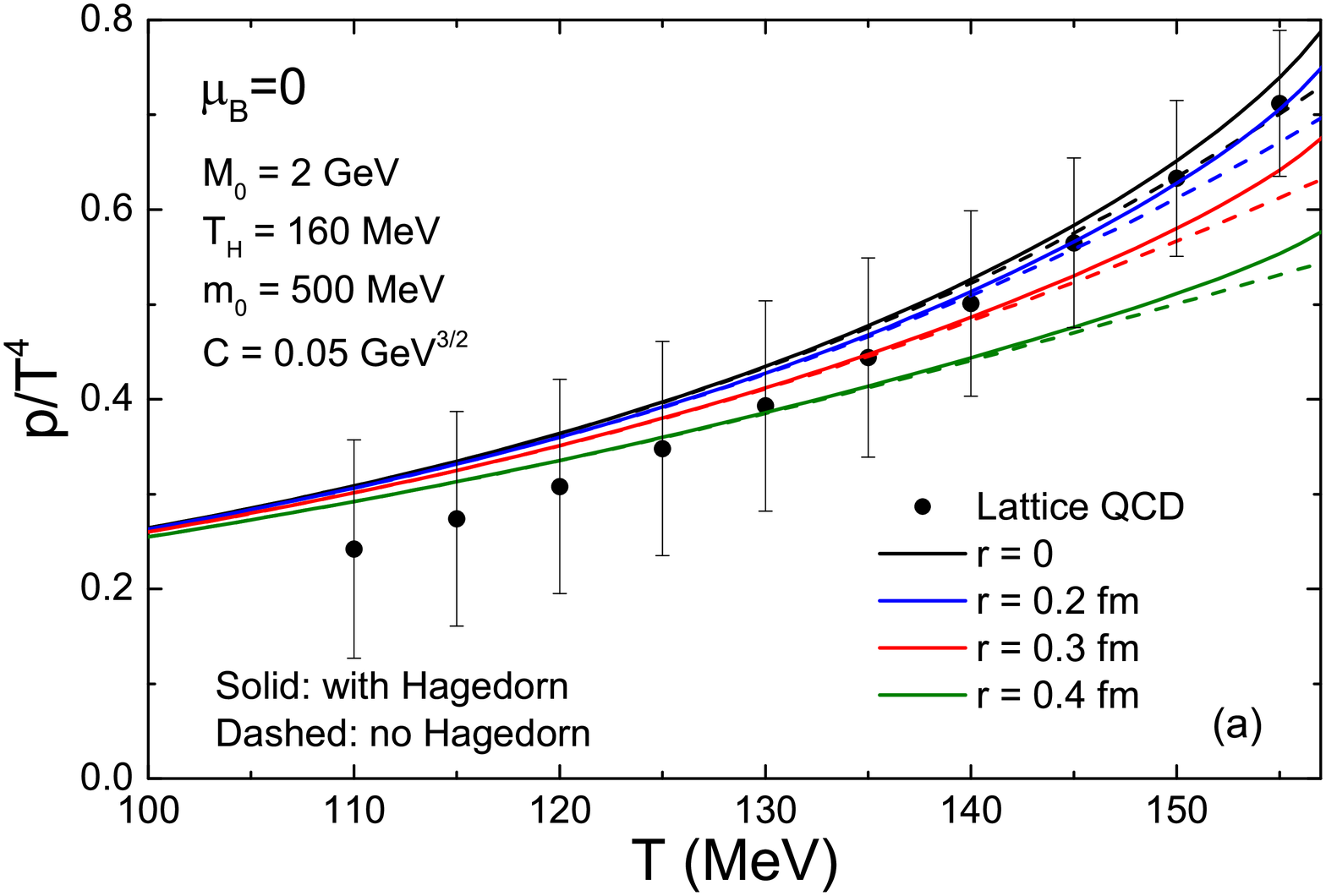}
\includegraphics[width=0.49\textwidth]{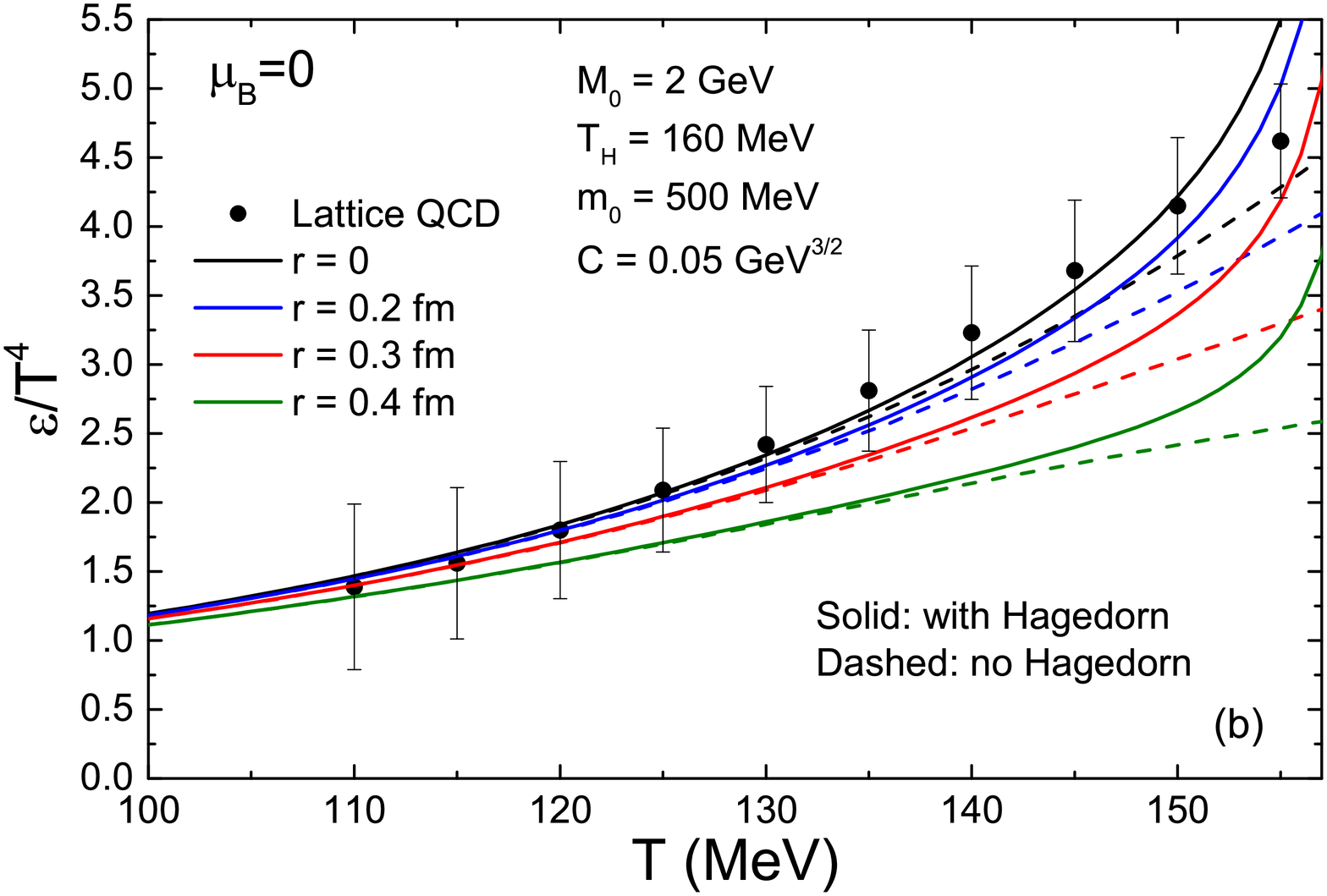}
\caption[]{The results of EV-HRG-H model for different values of $r$
are compared to the lattice data for $p/T^4$ and $\varepsilon/T^4$ in
(a) and (b), respectively.
The value of $C$ is fixed as $C = 0.05$~GeV$^{3/2}$.
}
\label{fig-hag}
\end{figure}

\begin{figure}[ht]
\centering
\includegraphics[width=0.49\textwidth]{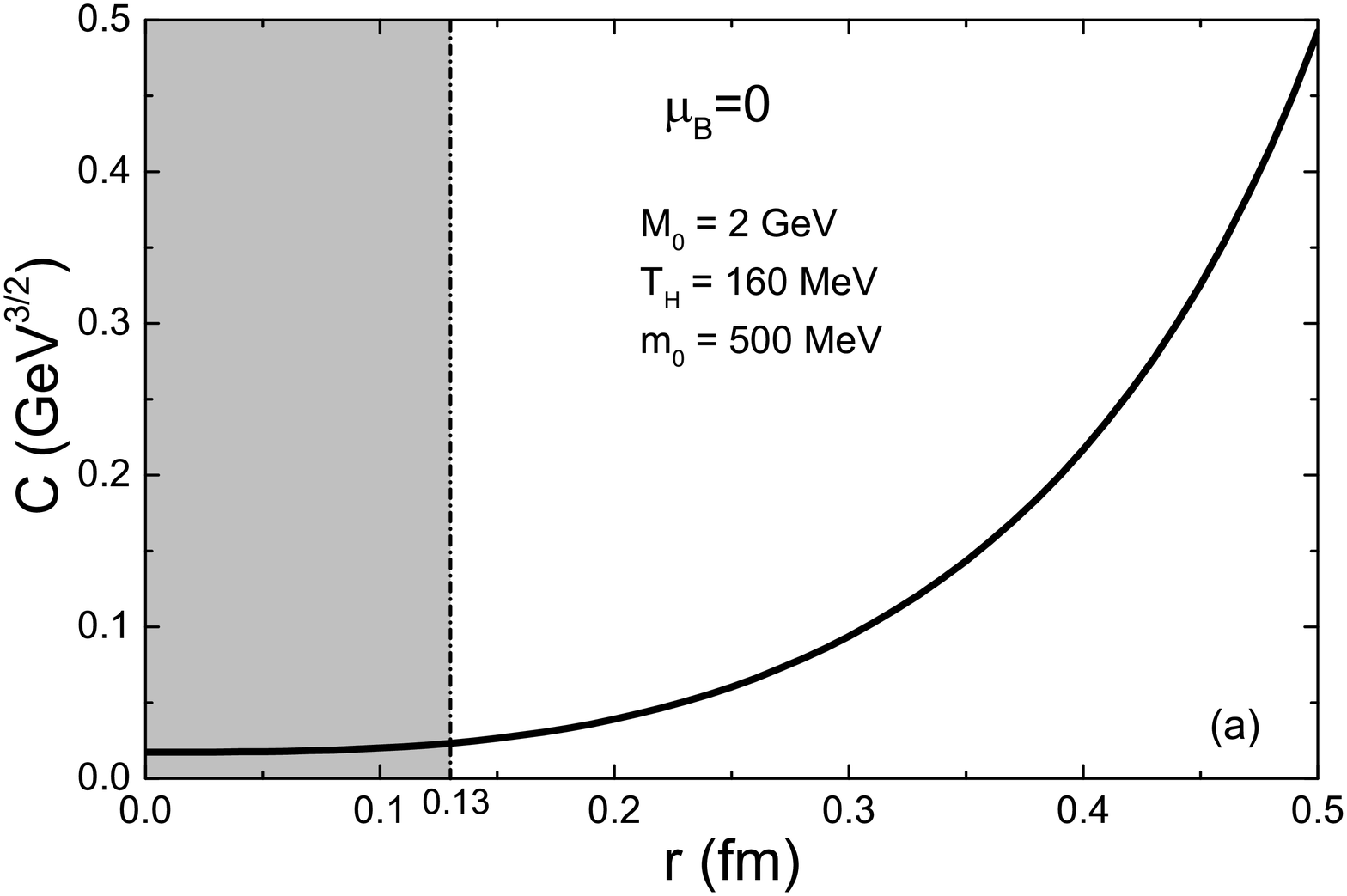}
\includegraphics[width=0.49\textwidth]{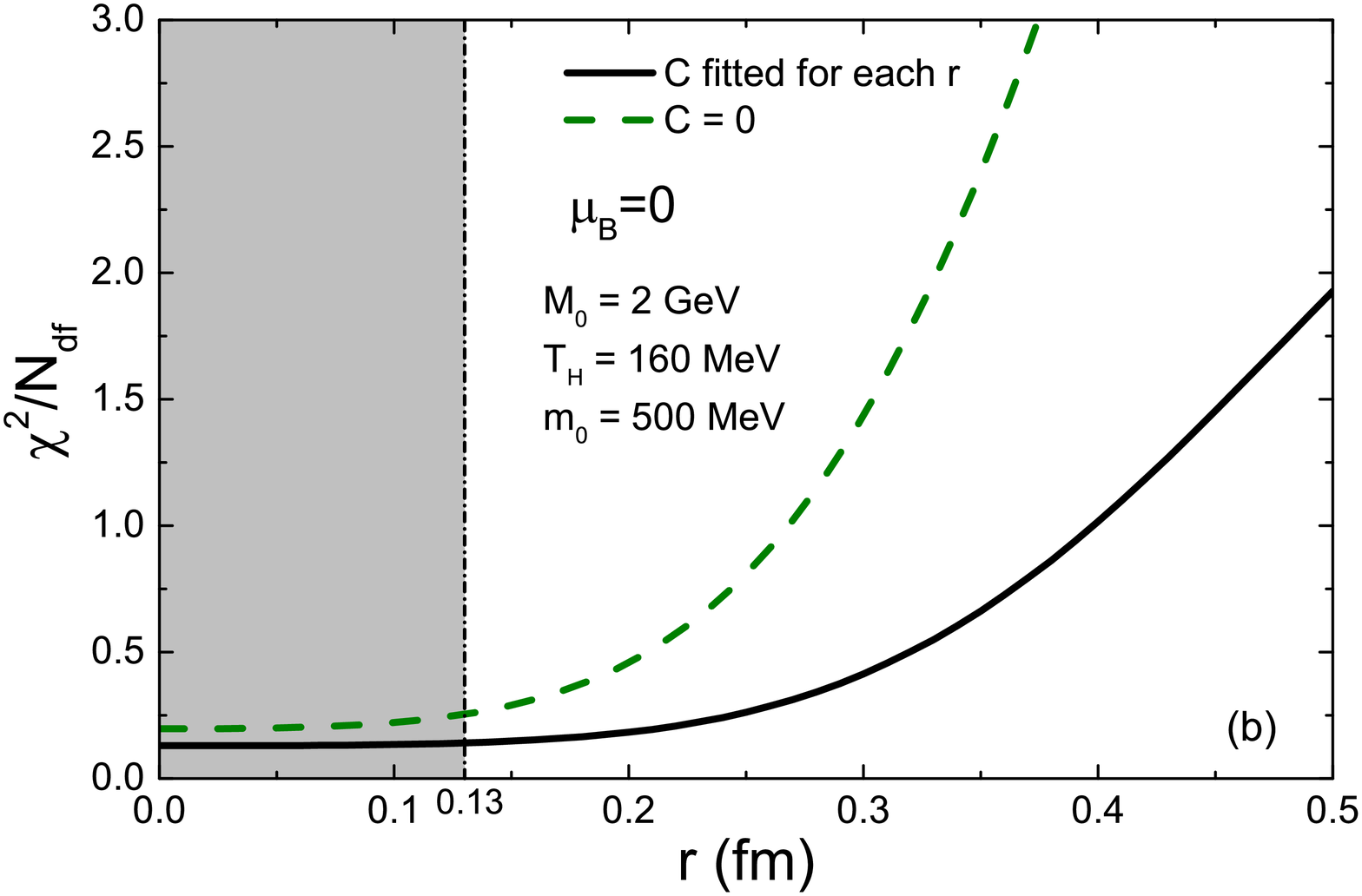}
\caption[]{
(a):
Parameter $C$  which minimizes
$\chi^2/N_{\rm df}$ at each value of $r$ is shown as a function of $r$.
(b): The quantity $\chi^2/N_{\rm df}$
as a function of $r$.
For each value of $r$ parameter $C$ is fitted
in order to minimize $\chi^2/N_{\rm df}$.
The shaded grey area corresponds to
$r\leq 0.13$.
}
\label{fig-hag2}
\end{figure}

In Fig.~\ref{fig-hag}, a comparison of the EV-HRG-H
with the
lattice data \cite{lattice-1c}
is presented.
The results for $p^{\rm H}/T^4$ and $\varepsilon^{\rm H}/T^4$ are presented by
the solid lines in Fig.~\ref{fig-hag} (a) and (b), respectively. These lines correspond
to different values of $r$ but fixed $C=0.05$~GeV$^{3/2}$.
The dashed lines
show the EV-HRG results without
Hagedorn mass spectrum, i.e., at
$C=0$. An inclusion  of the  Hagedorn mass spectrum become clearly visible at $T>130$~MeV,
and its contribution to the energy density is essentially larger than that to the pressure.

A simultaneous fit of the $p^{\rm lat}/T^4$ and $\varepsilon^{\rm lat}/T^4$
lattice data is done.
A quality of the fit is now controlled by $\chi^2/N_{\rm df}$
with $20=10+10$ number of points and 2 fitting parameters ($r$ and $C$).
At each value of $r$ one can find the $C$ parameter which minimize $\chi^2/N_{\rm df}$
at fixed $r$.
This introduces  the correlation
between parameters $C$ and $r$ which is shown in Fig.~\ref{fig-hag2} (a).
The data are well fitted for a rather wide
range of values for hard-core radius: $\chi^2/N_{\rm df} \lesssim 1$
for $r \lesssim 0.4$~fm.
In Fig.~\ref{fig-hag2} (b)
a dependence of $\chi^2/N_{\rm df}$ on $r$ is shown for $C=0$ and $C=C(r)$,
where $C(r)$ is depicted in
Fig~\ref{fig-hag2} (a).
A simultaneous fit of $p^{\rm lat}/T^4$ and $\varepsilon^{\rm lat}/T^4$ within EV-HRG model
(i.e., for $C=0$) does not show a necessity of $r>0$. Similarly,
the Id-HRG-H model (i.e., at $r=0$) admits only very small contributions from Hagedorn states
to the thermodynamical functions (a small value of  $C$ at $r=0$ seen in Fig.~\ref{fig-hag2} (a)).
Therefore, no clear evidence for $r>0$ or $C>0$ can be found if these
two effects are considered separately, i.e., within EV-HRG or Id-HRG-H model.
On the other hand, taking them simultaneously within EV-HRG-H model
we indicate a presence of these two effects and improvement of
the Id-HRG model.

Let us also note that an inclusion of the Hagedorn mass spectrum led to a successful
description of
lattice data for the confined glueball phase in the pure SU(3) case (without quarks)
in Ref.~\cite{Meyer}. This analysis is based on the assumption that glueball decay widths 
and interactions (i.e., excluded volume effects too) are rather small and can be neglected.
In the full SU(3) theory (with quarks) this assumption may not be valid.

\section{Summary}
\label{sec-sum}

In summary, the lattice data of Ref.~\cite{lattice-1c} for $p^{\rm lat}/T^4$ and
$\varepsilon^{\rm lat}/T^4$
are considered with the HRG model.
Two extensions of this model are analyzed: the excluded volume effects (with the
same hard-core radius $r$ for all particles) and the exponential Hagedorn mass
spectrum.
A condition that the pressure of the HRG should not exceed the Stefan-Boltzmann limit
for quarks and gluons indicates that hadrons should have
a non-zero hard-core radius of at least $0.13$~fm.
However, a comparison of the excluded volume HRG model with the lattice data at $T<155$~MeV yields
no
conclusive evidences in favor of a presence of the excluded volume effects.
Namely, the fit of  $p^{\rm lat}/T^4$ prefers
values of $r\lesssim 0.4$~fm,
while the best fit of $\varepsilon^{\rm lat}/T^4$ corresponds to $r\cong 0$.
If $r=0$, there is also not much room for the contribution from the Hagedorn
states, the best fit in this case corresponds to $C\cong 0$,
i.e. it suggests an absence of the contributions from the Hagedorn states.

It means that neither excluded volume HRG
nor ideal HRG with additional Hagedorn states being considered separately
demonstrates any advantages
for fitting the lattice data in a comparison to the ideal HRG model
(with no excluded volume effects and no Hagedron states).
On the other hand, if both these physical effects are considered simultaneously
the situation
is changed:
the data are well fitted for $r \lesssim 0.4$~fm and $C\lesssim 0.2$~GeV$^{3/2}$
with $\chi^2/N_{\rm df} \lesssim 1$,
i.e., there is a clear indication that both the hard core repulsion and
the Hagedorn mass spectrum should be taken into account simultaneously in the
framework of the hadron resonance gas model.
They
lead to rather essential contributions: suppression effects for
$p^{\rm H}/T^4$ and $\varepsilon^{\rm H}/T^4$  due to the excluded
volume effects and  enhancement due to the Hagedorn mass spectrum.
These {\it simultaneous} contributions ensure
a better agreement with the lattice data
and  lead, therefore, to
improvement of the ideal HRG model.

\begin{acknowledgments}
%
We would like to thank Z.~Fodor,
M.~Ga\'zdzicki, F.~Karsch, K.~Redlich, and E.~Shuryak
for fruitful comments.
Publication is based on the research provided by the grant support of the State 
Fund for Fundamental Research (project No. F58/175-2014). The work was partially supported 
by HIC for FAIR within the LOEWE program of the State of Hesse and by the Program of Fundamental Research of the Department of Physics and Astronomy of NAS of Ukraine (project No. 0112U000056).
\end{acknowledgments}

\end{document}